\documentclass[aps,prb,showpacs,twocolumn,floats,epsfig,pdflatex, superscriptaddress]{revtex4-2}
\usepackage{amssymb}
\usepackage{amsbsy}
\usepackage{amsmath}
\usepackage{epsfig}
\usepackage{amsfonts}
\usepackage{graphicx}
\usepackage{amssymb}
\usepackage{upgreek}
\usepackage{dsfont}
\usepackage{bm}
\usepackage{xcolor, comment}
\usepackage{hyperref}
\usepackage{tikz}

\begin{document}

\title {Detection of phonon helicity in nonchiral crystals with Raman scattering}

\author{Selçuk Parlak}
\affiliation{D\'epartement de Physique, Institut Quantique and Regroupement Qu\'eb\'ecois sur les Mat\'eriaux de Pointe, Universit\'e de Sherbrooke, Sherbrooke, Qu\'ebec, Canada J1K 2R1}
\author{Sayandip Ghosh}
\affiliation{D\'epartement de Physique, Institut Quantique and Regroupement Qu\'eb\'ecois sur les Mat\'eriaux de Pointe, Universit\'e de Sherbrooke, Sherbrooke, Qu\'ebec, Canada J1K 2R1}
\affiliation{Department of Physics, Visvesvaraya National Institute of Technology Nagpur, Nagpur 440010, India}
\author{Ion Garate}
\affiliation{D\'epartement de Physique, Institut Quantique and Regroupement Qu\'eb\'ecois sur les Mat\'eriaux de Pointe, Universit\'e de Sherbrooke, Sherbrooke, Qu\'ebec, Canada J1K 2R1}
\date{\today}

\begin{abstract}
Recently, it has been predicted that the Berry curvature of electrons can produce an angular momentum for phonons.
In systems with time-reversal symmetry, the direction of the phonon angular momentum is locked to the phonon wave vector.
Accordingly, this phenomenon has received the name of ``phonon helicity".
Here, we present a theory to unveil the signatures of such phonon helicity using Raman scattering. 
We show that the intensity of Raman scattering for circularly polarized light in BaMnSb$_2$ (a prototypical nonchiral Dirac insulator) changes under a reversal of the phonon wave vector, and that the phonon helicity can be inferred from that change. 
 We compare our results to recent reports of Raman-based detection of phonon angular momentum in chiral crystals. 
\end{abstract}

\maketitle

\section{Introduction}



The Berry phase of electrons is a central concept in modern materials science  \cite{xiao2010berry, vanderbilt2018berry, moessner2021topological}.
In addition, over the past decade, the Berry phases of nonelectronic quasiparticles such as phonons \cite{liu2020topological} and magnons \cite{mcclarty2022topological} have received extensive attention.
In comparison, less is known about the question of how the Berry phase of one type of quasiparticle manifests itself in the behavior of another type of quasiparticle.

In recent years, several theoretical studies have made progress along the aforementioned direction by unveiling signatures of the electronic Berry curvature in the phonon dispersion \cite{barkeshli2012dissipationless, song2016detecting, rinkel2017signatures, chernodub2019chiral, rinkel2019influence, sengupta2020phonon, antebi2021anomaly, sukhachov2021anomalous}. 
Most recently, it has been predicted that the electronic Berry curvature can also affect the phonon polarization and even induce a phonon angular momentum \cite{hu2021phonon, liu2022probing, saparov2022lattice}. 



\par

In crystals with time-reversal symmetry, the phonon angular momentum induced by the electronic Berry curvature is referred to as ``phonon helicity", 
since the direction of the phonon angular momentum is locked to the direction of the phonon propagation.
The underlying mechanism for this phenomenon is a hybridization between two linearly polarized, orthogonal phonon modes.
The hybridization, caused by the electron-phonon interaction, has a nonzero imaginary part inherited from the electronic Berry curvature.
As a result, the hybridized phonons become elliptically polarized with a nonzero angular momentum.
 This effect, first predicted in Ref.~[\onlinecite{hu2021phonon}], awaits an experimental observation. 

Part of the experimental challenge resides in the fact that, due to time reversal symmetry, the total phonon angular momentum (which involves a sum over all phonon wave vectors) vanishes. 
In Ref.~[\onlinecite{hu2021phonon}], it was mentioned that a temperature gradient can be used to produce a nonzero total phonon angular momentum by breaking time-reversal symmetry, and that one can afterwards measure the total phonon angular momentum via the Einstein-de Haas effect \cite{hamada2018phonon}.
Yet, this is an indirect detection scheme that moreover requires driving the system out of equilibrium.
It would be desirable to have an experimental method that can access the {\em equilibrium} phonon helicity {\em locally} in momentum space.


The purpose of the present work is to give a theoretical proposition for the detection of the phonon helicity, based on circularly polarized Raman spectroscopy.
Following Ref.~[\onlinecite{hu2021phonon}], we focus our study on BaMnSb$_2$, which is a quasi two dimensional Dirac insulator.
Although the system of interest in this paper is BaMnSb$_2$, the underlying ideas can be more generally applicable.
In Sec.~\ref{sec:proposal}, we review the key results of phonon helicity in BaMnSb$_2$ and present a detection scheme based on Raman scattering.
In Sec.~\ref{sec:disc}, we compare our study with recent proposals on the detection of phonon angular momentum in chiral crystals.
Finally, we summarize our findings in Sec.~\ref{sec:conc}.
Appendix A contains a self-contained derivation of phonon helicity in BaMnSb$_2$, following an approach that is complementary to the one developed in Ref.~[\onlinecite{hu2021phonon}].
 Appendix B discusses some general relations for the elements of the Raman tensor.
Appendix C contains an example on how linearly polarized light is insufficient to detect the phonon helicity in BaMnSb$_2$.



\section{Signatures of phonon helicity in Raman scattering}
\label{sec:proposal}
\subsection{Phonon helicity in BaMnSb$_2$}

We begin by reviewing a few key aspects of BaMnSb$_2$. This is a layered material with a quasi-2D electronic structure. Near the Fermi level, the electronic dispersion hosts two 2D Dirac fermions of mass $m$ and $-m$, respectively \cite{liu2021spin}. These two Dirac fermions are related to one another by time-reversal symmetry. 
Because these electronic states originate mainly from Sb layers, they couple significantly to the vibrations of Sb atoms.

For a vanishing wave vector, the phonon-modes associated to in-plane ($xy$)  vibrations of Sb atoms have $A_1$ and $B_1$ symmetries. 
These are linearly polarized modes with polarization along $x$ and $y$, respectively.
Their frequencies at zero wave vector will be denoted as $\omega_A$ and $\omega_B$, with  $\omega_A\neq \omega_B$
since the crystal does not have a $C_{4z}$ symmetry.

The aforementioned phonons couple to the 2D massive Dirac fermions as pseudo-gauge fields \cite{hu2021phonon}. 
As a consequence, the phonon effective action for optical modes $A_1$ and $B_1$ acquires a Chern-Simons term and reads (see App.~\ref{sec:app} below)
\begin{equation}
\label{eq:S_ph}
S_{\rm ph} \propto \int_{{\bf q},\omega} \left(u_A^*, u_B^*\right) \left(\begin{array}{cc} \omega^2-\omega_A^2  & \Sigma_{AB} \\ \Sigma_{BA} & \omega^2 -\omega_B^2 \end{array}\right) \left(\begin{array}{c} u_A \\u_B\end{array}\right),
\end{equation}
where ${\bf q}=(q_x,q_y)$ is the phonon momentum, $u_A({\bf q},\omega)$ and $u_B({\bf q},\omega)$ are the normal coordinates corresponding to the $A_1$ and $B_1$ modes, respectively, and $\Sigma_{AB}$ is a hybridization term originating from electron-phonon coupling.
In general, $\Sigma_{AB}$ is a function of both ${\bf q}$ and $\omega$, with $\Sigma_{BA}({\bf q}, \omega) = \Sigma_{AB}(-{\bf q},-\omega)$.
Then, the phonon eigenfrequencies in the presence of hybridization are obtained from the zeros of the determinant of the $2\times 2$ matrix in Eq.~(\ref{eq:S_ph}).
Also, the polarization vectors of the hybridized phonons are given by the eigenvectors of the $2\times 2$ matrix in Eq.~(\ref{eq:S_ph}), with $\Sigma_{AB}$  evaluated at the corresponding phonon eigenfrequency.

A microscopic calculation of $\Sigma_{AB}$ in the long-wavelength regime has been carried out in Ref.~[\onlinecite{hu2021phonon}] (see also App.~\ref{sec:app}  below for an alternative but equivalent derivation). Here, we summarize the most important features.
The real part of $\Sigma_{AB}$ vanishes if the following conditions are simultaneously realized: (1) zero temperature, (2) no disorder, (3) phonon frequency smaller than the energy gap of the insulator $2 |m|$.
Otherwise, in the most generic situation, $\Sigma_{AB}$ has both real and imaginary parts.
The $C_{2x}$ symmetry operation in BaMnSb$_2$ imposes\cite{hu2021phonon}
$\Sigma_{A B}(q_x,q_y, \omega) = -\Sigma_{AB}(q_x,-q_y,\omega)$.
Consequently, at long wavelength (small $q_x$ and $q_y$), both real and imaginary parts of $\Sigma_{AB}$ are proportional to $q_y$.
In particular, the static imaginary part
\begin{equation}
{\rm Im}\Sigma_{AB}(q_x, q_y,0) \propto  {\rm sgn}(m) q_y,
\end{equation}
 associated to the Chern-Simons term in the action, is remarkable in that (1) it originates from the Berry curvature (more precisely, ${\rm sgn}(m)$ is the valley Chern number of electrons), (2) it provides a complex phase to the hybridization between the $A_1$ and $B_1$ modes, thereby producing elliptically polarized phonon modes with a nonzero angular momentum along the $z$ axis, (3)  it is proportional to $q_y$ at long wavelength. 
Because of the latter property, the phonon angular momentum is reversed under $q_y\to -q_y$, and the hybridized phonons are  ``helical".

 The reason why $q_y\neq 0$ is required for a phonon angular momentum in BaMnSb$_2$ can be understood intuitively from symmetry arguments. When $q_y=0$, the  $C_{2x}$ axis and the $xz$ mirror plane of the crystal prevent any phonon angular momentum because the $y$ and $x$ components of the angular momentum would change sign under the $C_{2x}$ axis and the $xz$ mirror, respectively, while the $z$ component of the angular momentum would change sign under either the $C_{2x}$ axis or the $xz$ mirror. 
When ${\bf q}= q \hat{\bf x}$, both the $C_{2x}$ axis and the $xz$ mirror are preserved in the little group. As such, there cannot be a phonon angular momentum in this case either.
Yet, when $q_y\neq 0$, both the $C_{2x}$ axis and the $xz$ mirror are effectively broken (absent in the little group). Hence, in this case a phonon angular momentum is allowed.
In order to understand why the only possible direction of the phonon angular momentum is the $z$ direction, we need to recall that the crystal also has a $xy$ mirror plane, which is preserved by an in-plane phonon momentum. Since the $x$ and $y$ components of a phonon angular momentum would be odd under the $xy$ mirror, it follows that such an angular momentum is forbidden by symmetry. In contrast, an angular momentum along the $z$ direction is invariant under the $xy$ mirror, which is why it is allowed.


\subsection{Raman tensors of helical phonons in BaMnSb$_2$}

In its simplest realization, Raman scattering describes the inelastic scattering of light off lattice vibrations: 
an incident light of wave vector ${\bf k}_i$ and frequency $\omega_i$ excites a phonon mode $\lambda$ wave vector ${\bf q}$ and frequency $\omega_{\lambda,{\bf q}}$, so that the outgoing light has a wave vector ${\bf k}_s={\bf k}_i -{\bf q}$ and a frequency $\omega_s = \omega_i - \omega_{\lambda,{\bf q}}$.
The intensity of the light reaching the detector can be written as
\begin{equation}
    I_\lambda \propto\left | \hat{\bf e}_i^\dagger\cdot {\bf R}_\lambda \cdot \hat{\bf e}_s\right |^2,
    \label{eq:mainprop}
\end{equation}
where 
$\hat{\bf e}_i$ and $\hat{\bf e}_s$ are the electric field directions for the light reaching the sample and the light reaching the detector, respectively, and ${\bf R}$ is the Raman tensor represented as a $3\times 3$ matrix. 
In order to satisfy Maxwell's equations in empty space, $\hat{\bf e}_i\cdot {\bf k}_i = 0 = \hat{\bf e}_s\cdot {\bf k}_s$. 
The proportionality factor in Eq.~(\ref{eq:mainprop}) is a function of $\omega_i$, $\omega_{\lambda, {\bf q}}$, the temperature and the sample volume~\cite{cardona1982resonance}. 

The form of the Raman tensor in BaMnSb$_2$ can be determined from group theory \cite{loudon2001raman}.
The point group of BaMnSb$_2$ is $C_{2v}$.
Accordingly, the Raman tensors for the ${\bf q}=0$  phonon modes $A_1$ and $B_1$ have the form
\begin{equation}
{\bf R}_{A} =\begin{pmatrix}
a &0 & 0 \\
0&b& 0 \\
0&0 & c
\end{pmatrix}; 
{\bf R}_{B} =\begin{pmatrix}
0&d & 0 \\
d&0 & 0 \\
0&0 & 0 
\end{pmatrix},
\label{eq:ramantensor}
\end{equation}
where $a$, $b$, $c$, $d$ are coefficients related to the intrinsic properties of the crystal~\footnote{The Raman tensors in Eq. (\ref{eq:ramantensor}) differ by convention from those of Ref. \cite{loudon2001raman}, because in our case the $C_2$ axis is along $x$ (as opposed to $z$).}.
In general, these coefficients are complex numbers, due for example to the nonzero imaginary part of the dielectric function \cite{strach1998determination}. Complex phases of the Raman coefficients have been recently observed in low-symmetry 2D materials \cite{ribeiro2015unusual, resende2020origin, han2022complex}. 
The symmetric matrix structure of ${\bf R}_B$ in Eq.~(\ref{eq:ramantensor}) is an approximation, justified when the effect of the phonon frequency in the Raman tensor elements is negligible, i.e. far from resonant conditions (see App. \ref{sec:appB}).


In the presence of $\Sigma_{AB}$, the $A_1$ and $B_1$ modes get mixed.
For pseudo-gauge field type of electron-phonon coupling, we have $\Sigma_{BA}=\Sigma_{AB}^*$ (see App. \ref{sec:app}).
From now on, we will restrict ourselves to this case. 
Accordingly, the $2\times 2$ matrix in  Eq.~(\ref{eq:S_ph}) is hermitian and the normal coordinates for the hybridized modes (hereafter denoted as $+$ and $-$) are
\begin{align}
\label{eq:eigenmodes}
\left(\begin{array}{c} u_+ \\ u_-\end{array}\right)= \left(\begin{array}{cc}\cos\frac{\theta_{\bf q}}{2} &  e^{i \varphi_{\bf q}} \sin\frac{\theta_{\bf q}}{2}\\
-\sin\frac{\theta_{\bf q}}{2} &  e^{i \varphi_{\bf q}} \cos\frac{\theta_{\bf q}}{2}\end{array}\right)\left(\begin{array}{c} u_A \\ u_B\end{array}\right),
\end{align}
where 
$\sin\theta_{\bf q} = |\Sigma_{AB}|/\sqrt{(\omega_B^2-\omega_A^2)^2 + |\Sigma_{AB}|^2}$ gives a measure of the magnitude of the hybridization and $\tan\varphi_{\bf q} = -{\rm Im}\Sigma_{AB}/{\rm Re}\Sigma_{AB}$ gives a measure of the handedness of the hybridization.
The phonon helicity is produced by ${\rm Im}\Sigma_{AB}$, as it entails a nonzero value of $\varphi_{\bf q}$.
A quantitative measure of helicity  is given by $\sin\theta_{\bf q} \sin\varphi_{\bf q}$:
$\pm 1$ corresponding to circularly polarized (maximally helical) phonons and $0$ corresponding to linearly polarized (non helical) phonons. 
In BaMnSb$_2$, $\omega_A\neq \omega_B$ and thus the hybridized phonons are elliptically polarized.
Since both ${\rm Re}\Sigma_{AB}$ and ${\rm Im}\Sigma_{AB}$ are odd functions of $q_y$, the transformation $q_y \to -q_y$ implies $\varphi_{\bf q} \to \varphi_{\bf q} + \pi$ (mod $2\pi$).
This property will be useful below.


The Raman tensors for the hybridized phonons are
\begin{align}
\label{eq:Rpm}
{\bf R}_+ &= \cos\left(\theta_{\bf q}/2\right) {\bf R}_{A} +e^{-i \varphi_{\bf q}} \sin\left(\theta_{\bf q}/2\right)  {\bf R}_{B} \nonumber\\
{\bf R}_- &= -\sin\left(\theta_{\bf q}/2\right){\bf R}_{A}  +  e^{-i \varphi_{\bf q}} \cos\left(\theta_{\bf q}/2\right) {\bf R}_{B}.
\end{align}
This type of linear combination of Raman tensors has been used to describe the splitting of $G$ phonons of graphene under strain 
 \cite{huang2009phonon}. 
Phenomenologically, Eq.~(\ref{eq:Rpm})  can be derived from the chain rule. For example,
\begin{align}
\label{eq:chain}
{\bf R}_+ &= \frac{\partial\boldsymbol{\chi}}{\partial u_+} =\frac{\partial\boldsymbol{\chi}} {\partial u_A}\frac{\partial u_A}{\partial u_+}+\frac{\partial\boldsymbol{\chi}} {\partial u_B}\frac{\partial u_B}{\partial u_+}\nonumber\\
&={\bf R}_A\frac{\partial u_A}{\partial u_+}+{\bf R}_B\frac{\partial u_B}{\partial u_+},
\end{align}
where $\boldsymbol{\chi}$ is the electric susceptibility tensor.
In Eq.~(\ref{eq:chain}), the coefficients $\partial u_A/\partial u_+$ and $\partial u_B/\partial u_+$ can be read out from inverting Eq.~(\ref{eq:eigenmodes}).
The outcome is the first line of Eq.~(\ref{eq:Rpm}).

From a microscopic viewpoint, Eq.~(\ref{eq:Rpm})  can be derived by recognizing that the electron-phonon interaction matrix element 
associated to the creation of a phonon 
is proportional to~\cite{rinkel2017signatures}
 \begin{equation}
 \label{eq:me}
\sum_s \frac{1}{M_s} e^{i{\bf q}\cdot{\bf t}_s} {\bf p}^*_{\lambda,{\bf q},s}\cdot {\bf U}_{{\bf q},s},
 \end{equation}
 where $s$ is the atom label in the unit cell, $M_s$ is its mass, ${\bf t}_s$ is its position with respect to the origin of the unit cell, ${\bf p}_{\lambda,{\bf q},s}$ is the polarization vector, and ${\bf U}$ contains the matrix element of the derivative of the interatomic potential between electronic Bloch states~\cite{giustino2017electron}.
Equation ~(\ref{eq:me}) enters directly in the calculation of the Raman scattering amplitude \cite{loudon1963theory}.
For the optical phonons of interest in BaMnSb$_2$, we have $s=1,2$ (denoting two Sb atoms) and ${\bf p}_{\lambda,{\bf q},1}=-{\bf p}_{\lambda,{\bf q},2}$. 
Recalling that the polarization vectors of the unhybridized $A_1$ and $B_1$ phonons are $\hat{\bf x}$ and $\hat{\bf y}$, respectively, we substitute
\begin{align}
\label{eq:p_pm}
{\bf p}_{+,{\bf q},s} &\propto \cos(\theta_{\bf q}/2) \hat{\bf x} + e^{i\varphi_{\bf q}} \sin(\theta_{\bf q}/2) \hat{\bf y}\nonumber\\
{\bf p}_{-,{\bf q},s} &\propto -\sin(\theta_{\bf q}/2) \hat{\bf x} + e^{i\varphi_{\bf q}} \cos(\theta_{\bf q}/2) \hat{\bf y}
\end{align}
in Eq.~(\ref{eq:me}). It follows that
${\bf R}_\pm$ can be written as a linear combination of ${\bf R}_A$ and ${\bf R}_B$ with the coefficients given  in Eq.~(\ref{eq:Rpm}).

\subsection{Detection of phonon hybridization and helicity in BaMnSb$_2$}

Next, we propose a scheme to detect phonon hybridization and helicity with Raman spectroscopy. 
To begin with, by selectively choosing the polarizations of the incoming and scattered lights, it is possible to find an experimental evidence of the magnitude of the hybridization. 
For example, when $\hat{\bf e}_i = (0,0,z)$ and $\hat{\bf e}_s = (0,y', z')$ we get
\begin{align}
\label{eq:pol1}
I_+&\propto |c z z'|^2 \cos^2\left(\theta_{\bf q}/2\right) \nonumber\\
I_- &\propto |c z z'|^2 \sin^2\left(\theta_{\bf q}/2\right).
\end{align}
Accordingly, in the absence of hybridization (when $\theta_{\bf  q} = 0$ or $\pi$), only one of the two modes is visible in Raman (the $+$ mode if $\omega_B>\omega_A$; the $-$ mode otherwise).
Hybridization can be turned on and off by varying $q_y$. For example, if $\omega_B>\omega_A$, $I_- \propto |\Sigma_{AB}|^2 \propto q_y^2$ at long wavelength. Concomitantly, $I_+$ will decrease quadratically with $q_y$.
The complementary situation is realized when $\hat{\bf e}_i = (x,0,z)$ and $\hat{\bf e}_s = (0,y', 0)$, with
\begin{align}
\label{eq:pol2}
I_+&\propto |d x y'|^2 \sin^2\left(\theta_{\bf q}/2\right)\nonumber\\
I_- &\propto |d x y'|^2 \cos^2\left(\theta_{\bf q}/2\right).
\end{align}
In this case, for $\omega_B>\omega_A$, $I_+ \propto |\Sigma_{AB}|^2 \propto q_y^2$.


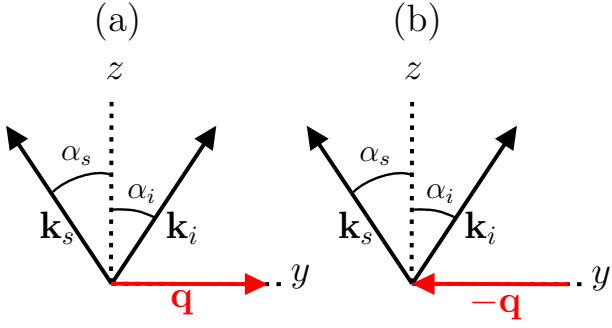
\begin{figure}
  \begin{center}

\tikzset{every picture/.style={line width=0.9pt}} 

\begin{tikzpicture}[x=0.75pt,y=0.75pt,yscale=-1,xscale=1]

\draw [line width=1.5]    (177.14,152.19) -- (126.53,228.67) ;
\draw [shift={(179.34,148.86)}, rotate = 123.49] [fill={rgb, 255:red, 0; green, 0; blue, 0 }  ][line width=0.08]  [draw opacity=0] (11.61,-5.58) -- (0,0) -- (11.61,5.58) -- cycle    ;
\draw [line width=1.5]    (75.56,152.19) -- (126.53,228.67) ;
\draw [shift={(73.34,148.86)}, rotate = 56.32] [fill={rgb, 255:red, 0; green, 0; blue, 0 }  ][line width=0.08]  [draw opacity=0] (11.61,-5.58) -- (0,0) -- (11.61,5.58) -- cycle    ;
\draw [line width=1.5]  [dash pattern={on 1.69pt off 2.76pt}]  (126.49,136.86) -- (126.2,228.67) ;
\draw    (96.91,182.35) .. controls (103.21,175.85) and (118.77,170.57) .. (126.2,173.71) ;
\draw [line width=1.5]  [dash pattern={on 1.69pt off 2.76pt}]  (211.91,228.86) -- (126.53,228.67) ;
\draw [color={rgb, 255:red, 255; green, 0; blue, 0 }  ,draw opacity=1 ][line width=1.5]    (201.34,228.85) -- (126.2,228.67) ;
\draw [shift={(205.34,228.86)}, rotate = 180.14] [fill={rgb, 255:red, 255; green, 0; blue, 0 }  ,fill opacity=1 ][line width=0.08]  [draw opacity=0] (11.61,-5.58) -- (0,0) -- (11.61,5.58) -- cycle    ;
\draw    (126.63,191.2) .. controls (138.49,190.86) and (140.77,192) .. (148.2,195.14) ;
\draw [line width=1.5]    (328.74,152.22) -- (278.13,228.7) ;
\draw [shift={(330.94,148.89)}, rotate = 123.49] [fill={rgb, 255:red, 0; green, 0; blue, 0 }  ][line width=0.08]  [draw opacity=0] (11.61,-5.58) -- (0,0) -- (11.61,5.58) -- cycle    ;
\draw [line width=1.5]    (227.16,152.21) -- (278.13,228.7) ;
\draw [shift={(224.94,148.89)}, rotate = 56.32] [fill={rgb, 255:red, 0; green, 0; blue, 0 }  ][line width=0.08]  [draw opacity=0] (11.61,-5.58) -- (0,0) -- (11.61,5.58) -- cycle    ;
\draw [line width=1.5]  [dash pattern={on 1.69pt off 2.76pt}]  (278.09,136.89) -- (277.8,228.7) ;
\draw    (248.51,182.37) .. controls (254.81,175.87) and (270.37,170.6) .. (277.8,173.74) ;
\draw [line width=1.5]  [dash pattern={on 1.69pt off 2.76pt}]  (363.51,228.89) -- (278.13,228.7) ;
\draw [color={rgb, 255:red, 255; green, 0; blue, 0 }  ,draw opacity=1 ][line width=1.5]    (356.94,228.89) -- (281.8,228.7) ;
\draw [shift={(277.8,228.7)}, rotate = 0.14] [fill={rgb, 255:red, 255; green, 0; blue, 0 }  ,fill opacity=1 ][line width=0.08]  [draw opacity=0] (11.61,-5.58) -- (0,0) -- (11.61,5.58) -- cycle    ;
\draw    (278.23,191.23) .. controls (290.09,190.89) and (292.37,192.03) .. (299.8,195.17) ;

\draw (102.42-3,157.93) node [anchor=north west][inner sep=0.75pt]   [align=left] {\large $\alpha_s$};
\draw (135.96-3,172.94+4) node [anchor=north west][inner sep=0.75pt]   [align=left] {\large $\alpha_i$};
\draw (154.61,225.18+5) node [anchor=north west][inner sep=0.75pt]   [align=left] {\textcolor[rgb]{1,0,0}{\Large ${\bf q}$}};
\draw (152.26,190.7) node [anchor=north west][inner sep=0.75pt]   [align=left] {\Large ${\bf k}_i$};
\draw (88.5,190.7) node [anchor=north west][inner sep=0.75pt]   [align=left] {\Large ${\bf k}_s$};
\draw (121.76-6,92-7) node [anchor=north west][inner sep=0.75pt]   [align=left] {\Large (a)};
\draw (121.76,115.54) node [anchor=north west][inner sep=0.75pt]   [align=left] {\Large $z$};
\draw (215.36,217.54) node [anchor=north west][inner sep=0.75pt]   [align=left] {\Large $y$};
\draw (228.02+25-3,157.96) node [anchor=north west][inner sep=0.75pt]   [align=left] {\large $\alpha_s$};
\draw (261.56+25-3,172.97+4) node [anchor=north west][inner sep=0.75pt]   [align=left] {\large $\alpha_i$};
\draw (280.21+25,225.21+5) node [anchor=north west][inner sep=0.75pt]   [align=left] {\textcolor[rgb]{1,0,0}{\Large ${\bf -q}$}};
\draw (277.86+25,190.73) node [anchor=north west][inner sep=0.75pt]   [align=left] {\Large ${\bf k}_i$};
\draw (214.1+25,190.73) node [anchor=north west][inner sep=0.75pt]   [align=left] {\Large ${\bf k}_s$};
\draw (247.36-6+25,92-7) node [anchor=north west][inner sep=0.75pt]   [align=left] {\Large (b)};
\draw (247.36+25,115.57) node [anchor=north west][inner sep=0.75pt]   [align=left] {\Large $z$};
\draw (341.96+25,218.57) node [anchor=north west][inner sep=0.75pt]   [align=left] {\Large $y$};

\end{tikzpicture}

\caption{Schematic setup showing incoming and outgoing light with wave vectors ${\bf k}_i$ and ${\bf k}_s$.
In the main text, we consider $\alpha_i=\alpha_s\equiv\alpha$ for simplicity, though the qualitative features of our main results do not rely on this assumption.
When $q_y\neq 0$, the excited phonon is elliptically polarized in the $xy$ plane and has an angular momentum along $z$. The ellipticity and angular momentum are reversed under $q_y\to -q_y$. 
As explained in the main text, the ellipticity of the phonon can be detected by using incident and scattered lights that have nonzero angular momentum along the $z$ axis ($\alpha\neq \pi/2$), while also keeping a nonzero momentum along $y$ ($\alpha\neq 0$). 
(a) Stokes process: emission of a phonon of wave vector ${\bf q}$; (b) anti-Stokes process: absorption of a phonon of wave vector $-{\bf q}$.
 }
\label{fig:expsketch}
\end{center}
\end{figure}

In Eqs.~(\ref{eq:pol1}) and (\ref{eq:pol2}), there is no information about the handedness of the phonon hybridization. Are the hybridized phonons linearly or elliptically polarized? The answer to this question is contained in $\varphi_{\bf q}$. 
Combining Eqs.~(\ref{eq:mainprop}) and (\ref{eq:Rpm}), it is easy to show that, for generic $\hat{\bf e}_i=(x,y,z)$ and $\hat{\bf e}_s=(x',y',z')$, we need $a x^* x' + b y^* y'+ c z^* z'\neq 0$ and $x^* y'+x' y^* \neq 0$ in order to access $\varphi_{\bf q}$ in the Raman intensity. This gives certain freedom of configuration.

For concreteness, we select the configuration displayed in Fig.~\ref{fig:expsketch}a, with $\alpha_s=\alpha_i\equiv\alpha$:
\begin{align}
{\bf k}_i &=\frac{\omega_i}{c_0}  \left(0, \sin\alpha, \cos\alpha\right) \notag\\
 {\bf k}_s &=\frac{\omega_s}{c_0}  \left(0, -\sin\alpha, \cos\alpha\right),
\end{align}
$c_0$ being the speed of light.
In this configuration, the phonon momentum is ${\bf q} \simeq \hat{\bf y} (2\omega_i/c_0) \sin\alpha$, assuming that $\omega_{\lambda, {\bf q}}\ll \omega_i$.
Moreover, we assume that the incident and scattered lights are circularly polarized:
\begin{align}
\label{eq:polvector}
\hat{\bf e}_i &=\frac{1}{\sqrt{2}} \left(1, \pm i \cos\alpha, \mp i \sin\alpha\right)\notag\\
\hat{\bf e}_s &=\frac{1}{\sqrt{2}}  \left(1, \pm i \cos\alpha, \pm i \sin\alpha\right), 
\end{align}
where the upper and lower signs correspond to right-($R$) and left-($L$) circularly polarized light. The appearance of the $\cos\alpha$ and $\sin\alpha$ factors in Eq.~(\ref{eq:polvector}) is due to the rotation of the coordinate system from the light's frame (where $z$ axis is defined by the propagation  direction) to the laboratory frame.
In App. \ref{sec:appC}, we show that linearly polarized light is not conducive to the detection of the phonon helicity.
 It makes intuitive sense that, in order for the Raman intensity to be sensitive to the phonon's angular momentum, the light must also have a nonzero angular momentum.

Combining Eqs.~(\ref{eq:mainprop}), (\ref{eq:Rpm}) and (\ref{eq:polvector}), it follows that the Raman intensity depends on $\varphi_{\bf q}$
only when the circularly polarized incident and scattered lights have opposite handedness: we denote these configurations  $LR$ and $RL$.
Then, Eq.~(\ref{eq:mainprop}) yields
\begin{align}
      I_+^{RL} &\propto \frac{1}{4} \cos^2 \left(\frac{\theta_{\bf q}}{2}\right) \left|f\right|^2+\sin^2 \left(\frac{\theta_{\bf q}}{2}\right) |g|^2 -I_{\rm int}\notag\\
       I_+^{LR} &\propto \frac{1}{4} \cos^2 \left(\frac{\theta_{\bf q}}{2}\right) \left|f\right|^2+\sin^2 \left(\frac{\theta_{\bf q}}{2}\right) |g|^2 +I_{\rm int}\notag\\
          I_-^{RL} &\propto \frac{1}{4} \sin^2 \left(\frac{\theta_{\bf q}}{2}\right) \left|f\right|^2+\cos^2 \left(\frac{\theta_{\bf q}}{2}\right) |g|^2 +I_{\rm int}\notag\\
       I_-^{LR} &\propto \frac{1}{4} \sin^2 \left(\frac{\theta_{\bf q}}{2}\right) \left|f\right|^2+\cos^2 \left(\frac{\theta_{\bf q}}{2}\right) |g|^2 -I_{\rm int}
    \label{eq:LRintensity}
\end{align}
where
$f \equiv  a-b \cos^2\alpha + c\sin^2\alpha$, $g \equiv d\cos\alpha$, 
 and
\begin{equation}
\label{eq:I_int}
I_{\rm int} = \frac{1}{2}|f g|\sin \theta_{\bf q} \sin(-\varphi_{\bf q}+\phi_g-\phi_f)
\end{equation}
is the interference term between the $A_1$ and $B_1$ modes.
In addition, $\phi_f$ and $\phi_g$ are the phases of the complex numbers $f$ and $g$, respectively. These phases are determined by ${\bf q}=0$ Raman matrix elements, as well as by $\alpha$. 
For small but nonzero phonon wave vectors, we anticipate that $\phi_g-\phi_f$ will change smoothly.
In contrast, $\varphi_{\bf q}$ has an abrupt, step-like dependence on $q_y$, since  ${\rm Re}\Sigma_{AB}\propto q_y$ and ${\rm Im}\Sigma_{AB}\propto q_y$ at long wavelength. 

Various properties of $I_{\rm int}$ are worth mentioning. 
First, it is the only $\varphi_{\bf q}$-dependent term in the Raman intensity. As such, it contains the information about the phonon helicity.
Second, $I_{\rm int}$ appears with opposite signs for $+$ and $-$ modes, as well as for $RL$ and $LR$ configurations.
 It makes intuitive sense that, for a fixed phonon angular momentum, reversing the angular momenta of the incoming and scattered light will result in a change of the Raman intensity; such is the meaning of $I_{\rm int}\neq 0$.
Third, for long wavelength phonons, $\sin\theta_{\bf q} \propto |q_y|$. Recalling that $\varphi_{\bf q} \to \varphi_{\bf q} + \pi$ when $q_y \to -q_y$, it follows that $I_{\rm int} \propto q_y$. Hence, $I_{\rm int}$ is an odd function of $q_y$. 
This property of $I_{\rm int}$ is a reflection of a phonon angular momentum that is inverted under $q_y\to -q_y$.
The fact that the Raman intensity remains invariant under the concatenated application of $q_y\to -q_y$ and $LR \to RL$ is a consequence of  the $C_{2x}$ axis of the crystal's point group.
 Likewise, the fact that the Raman intensity remains invariant under the simultaneous transformations $LR \to RL$, $\varphi_{\bf q}\to - \varphi_{\bf q}$ and $\phi_f-\phi_g \to -(\phi_f-\phi_g)$ is a consequence of time reversal symmetry.
Another important aspect of $I_{\rm int}$ is that it vanishes when $\cos\alpha = 0$ (i.e. $g=0$); this corresponds to the configuration where the incident and scattered lights are along the $y$ direction. Thus, we learn that in order to have a difference in the Raman intensity between $RL$ and $LR$ configurations, the incoming and scattered light must have nonzero angular momentum along $z$, which is also the direction of the phonon angular momentum.
Also, since $|\Sigma_{AB}|\propto |q_y| \propto |\sin\alpha|$ and $|g| \propto \cos\alpha$, $\alpha=\pi/4$ gives the largest value for $I_{\rm int}$.

In order to experimentally single-out the phonon helicity, one can measure the difference in Raman intensities between $RL$ and $LR$ configurations: 
\begin{equation}
\label{eq:ID1}
I^{RL}_\pm-I^{LR}_\pm \propto \mp  I_{\rm int}.
\end{equation}
Alternatively, we can keep the photon polarizations fixed and compare the intensities for ${\bf q}=q_y\hat{\bf y}$ and ${\bf q}=-q_y\hat{\bf y}$:
\begin{align}
\label{eq:ID2}
&I^{RL}_\pm (q_y)-I^{RL}_\pm(-q_y) \propto \mp  I_{\rm int}\notag\\
&I^{LR}_\pm (q_y)-I^{LR}_\pm(-q_y) \propto \pm  I_{\rm int}.
\end{align}
In Eqs.~(\ref{eq:ID1}) and (\ref{eq:ID2}), it is assumed that the unwritten proportionality factor of Eq.~(\ref{eq:mainprop}) is {\em unchanged} under $LR \to RL$ and $q_y \to -q_y$, respectively.
For Eq.~(\ref{eq:ID1}), this assumption is correct  under the condition that birefringence effects make a negligible contribution~\cite{oishi2022selective}.
For Eq.~(\ref{eq:ID2}), the assumption is correct, as both the phonon frequency and the light refractive index are invariant under $q_y \to -q_y$ due to symmetry of BaMnSb$_2$.

From a quantitative standpoint, one can establish the relative size of $I_{\rm int}$ by taking the ratio between the difference and the sum of intensities in the $RL$ and $LR$ configurations. 
For weak hybridization ($|\Sigma_{AB}|^2\ll (\omega_A-\omega_B)^2$), we find
\begin{equation}
\label{eq:ratio}
\left|\frac{I^{RL}_+-I^{LR}_+}{I^{RL}_++I^{LR}_+}\right| \sim \frac{|g|}{|f|} \frac{|\Sigma_{AB}|}{|\omega_A-\omega_B|}.
\end{equation}
While Eq.~(\ref{eq:ratio}) appears potentially measurable, a more precise quantitative analysis goes beyond the scope of the present work. 

It is important to clarify that the experimental observation of the sign reversal of $I_{\rm int}$ under $q_y\to -q_y$ or under $LR \to RL$  does {\em not} prove  that the hybridized phonons are elliptically polarized. 
For instance, in the hypothetical case of  ${\rm Im}\Sigma_{AB}=0$ and ${\rm Re}\Sigma_{AB}\propto q_y \neq 0$  (which does not apply to BaMnSb$_2$), the hybridized phonons would be linearly polarized with $\varphi_{\bf q}=0$ or $\pi$. Yet, even in that case, $I_{\rm int}$  would reverse sign under $q_y\to -q_y$.

The origin of the difficulty lies at the phase $\phi_g-\phi_f$, which always appears together with $\varphi_{\bf q}$ in Eq.~(\ref{eq:I_int}). Should $\phi_g-\phi_f$ be zero or a multiple of $\pi$, then $I_{\rm int}\neq 0$ would be an automatic proof that the phonons are elliptically polarized and thus helical.
Yet, there is no reason that $\phi_g-\phi_f$ should be zero or a multiple of $\pi$.

Hence, to experimentally conclude that $\varphi_{\bf q}$ is not a multiple of $\pi$, additional information is needed.
At first sight, it is not obvious how to experimentally distinguish $\phi_g-\phi_f$ from $\varphi_{\bf q}$, as both occur together in the Raman intensity.
We propose an approach that consists of comparing the Raman intensities in the Stokes and anti-Stokes configuration (see Fig.~\ref{fig:expsketch}). 
Going from one to the other, the emission of a phonon of wave vector ${\bf q}$ is replaced by the absorption of a phonon of wave vector $-{\bf q}$. 
What does this imply for the phonon polarization? In order to answer this question, we recognize that the displacement operator for atom $s$ in unit cell ${\bf l}$, which appears in the matrix element of the electron-phonon interaction, is hermitian and can be written as \cite{rinkel2017signatures}
\begin{equation}
\label{eq:w}
\hat{\bf w}_{{\bf l}, s} \propto\sum_{\bf q}  \frac{e^{i {\bf q}\cdot{\bf l}}}{\sqrt{\omega_{\lambda, {\bf q}}}} \left({\bf p}_{\lambda, {\bf q}, s}^* \hat{a}^\dagger_{\lambda, {\bf q}} + {\bf p}_{\lambda, -{\bf q}, s} \hat{a}_{\lambda, -{\bf q}}\right).
\end{equation}
Here, $\hat{a}$ and $\hat{a}^\dagger$ are the phonon destruction and creation operators, and $\omega_{\lambda, {\bf q}} = \omega_{\lambda, -{\bf q}}$.
The first and the second terms in the right hand side of Eq.~(\ref{eq:w}) are associated to Stokes and anti-Stokes processes, respectively. 
Therefore, going from Stokes to anti-Stokes, the phonon polarization changes  from ${\bf p}_{\lambda {\bf q} s}^*$ to ${\bf p}_{\lambda,-{\bf q},s}$~\footnote{In the literature, one often finds the statement ${\bf p}_{\lambda, -{\bf q}, s} = {\bf p}_{\lambda, {\bf q}, s}^*$. Yet, this statement is not compatible with the polarization vectors of helical phonons in Eq.~(\ref{eq:p_pm}). Thus, we write Eq.~(\ref{eq:w}) in a more general way without assuming  ${\bf p}_{\lambda, -{\bf q}, s} = {\bf p}_{\lambda, {\bf q}, s}^*$}.
This entails $\exp(-i \varphi_{\bf q})\to\exp[i (\varphi_{\bf q}+\pi)]$ with $\theta_{\bf q}$ unchanged in Eq.~(\ref{eq:Rpm}).
On the other hand, the values of $|f|$, $|g|$ and $\phi_f-\phi_g$ are approximately the same for the Stokes to anti-Stokes processes, 
assuming that we are not close to resonance conditions \cite{loudon1963theory}. 
As a result, the ratio of $I_{\rm int}$ between the Stokes (S) and anti-Stokes (AS) configurations is
\begin{equation}
\frac{I_{\rm int}(\text{S})}{I_{\rm int}(\text{AS})}=-\frac{n(\omega_{\lambda,{\bf q}})+1}{n(\omega_{\lambda,{\bf q}})} \frac{\sin(-\varphi_{\bf q}+\phi_g-\phi_{f})}{\sin(\varphi_{\bf q}+\phi_g-\phi_{f})},
\label{eq:stokesanti}
\end{equation} 
where $n(\omega_{\lambda, {\bf q}})$ is the Bose-Einstein distribution for frequency $\omega_{\lambda,{\bf q}}$ and we have assumed that $\omega_i\gg \omega_{\lambda,{\bf q}}$.
If the hybridized phonon were linearly polarized $(\varphi_{\bf q} = 0,\pi$), the ratio in Eq.~(\ref{eq:stokesanti}) would take the  value $-[n(\omega_{\lambda, {\bf q}})+1]/n(\omega_{\lambda, {\bf q}})$. This ratio would be modified if the phonon were elliptically polarized $(\varphi_{\bf q} \neq  0,\pi$), thereby allowing to detect the phonon helicity.
For the case of maximal helicity ($\varphi_{\bf q}=\pi/2$), the ratio would be $+[n(\omega_{\lambda, {\bf q}})+1]/n(\omega_{\lambda, {\bf q}})$.

\section{Discussion}
\label{sec:disc}

In this section, we compare our work to earlier works in a related subject and lay down some important aspects of our findings. 
In particular, we focus on 
a preprint~\cite{oishi2022selective}, which appeared while our work was in preparation.
This preprint studies the detection of phonon angular momentum in a chiral crystal  ($\alpha-$quartz) using Raman scattering (see also Ref.~[\onlinecite{ishito2022truly}] for related work). 
There are two significant similarities between Ref.~[\onlinecite{oishi2022selective}] and our paper: (1) the idea of using $LR$ and $RL$ Raman configurations in order to detect the phonon angular momentum, and (2) the phenomenological reason for the emergence of a phonon angular momentum, i.e., a phonon hybridization that is linear in the phonon wave vector. 
In spite of these similarities, there are several differences between Ref.~[\onlinecite{oishi2022selective}] and our work, which are worth listing and discussing. 

First, Ref.~[\onlinecite{oishi2022selective}] emphasizes the importance of having a chiral crystal (such as $\alpha-$quartz) in order to have long-wavelength phonons with circular polarization. 
Our work evidences that the phonon helicity is a more general phenomenon that arises also in nonchiral crystals (such as BaMnSb$_2$), and that Raman scattering can be a useful tool for those more general systems as well.

Second, in $\alpha-$quartz, the $E$ phonons acquire an angular momentum when their double degeneracy at ${\bf q}=0$ is split for $q_z\neq 0$. Because the hybridization at $q_z\neq 0$ is stated to be purely imaginary, the hybridized phonons have perfectly circular polarization. 
In BaMnSb$_2$, the hybridization takes place between $A_1$ and $B_1$ phonons, which are nondegenerate at ${\bf q}=0$. In addition, the hybridization self-energy $\Sigma_{AB}$ can in general have both real and imaginary components.
As a result, the hybridized phonons in BaMnSb$_2$ are elliptically polarized. 

Third, in $\alpha-$quartz, only one of the two hybridized phonons is observable in the $LR$ or $RL$ configuration.
The Raman selection rules in BaMnSb$_2$ being different, both hybridized phonons can be observed in $LR$ and $RL$ configurations. 

Fourth, in  $\alpha-$quartz, reversing the sign of $q_z$ for a fixed Raman configuration (e.g. $LR$), or alternatively reversing the Raman configuration $LR \to RL$ for a fixed ${\bf q}$,  implies a change in the frequency of the observed phonon.
However, the value of  $|\hat{\bf e}_i^\dagger\cdot {\bf R}_\lambda \cdot \hat{\bf e}_s |^2$ is nearly unchanged (as the actual value of  $\omega_{\lambda,{\bf q}}$ matters little for ${\bf R}_\lambda$ away from resonance conditions~\cite{loudon1963theory}).
In contrast, in BaMnSb$_2$, reversing $q_y\to -q_y$  or  $LR \to RL$ leads to the {\em same} frequency of the observed phonon, albeit with a {\em different} value of $|\hat{\bf e}_i^\dagger\cdot {\bf R}_\lambda \cdot \hat{\bf e}_s |^2$ and thus of the Raman intensity.

Fifth, in $\alpha-$quartz, the direction of the phonon angular momentum ($z$) coincides with the direction of the phonon wave vector ($q_z$) at which the phonon hybridization is realized. As a result, the backscattering Raman configuration (${\bf k}_i || \hat{\bf z}$ and ${\bf k}_s || -\hat{\bf z}$) is ideally suited to detect the phonon angular momentum. 
In BaMnSb$_2$, the direction of the phonon angular momentum ($z$) is {\em perpendicular} to  the phonon wave vector ($q_y$) at which the phonon hybridization is realized. As a result, the backscattering configuration  (${\bf k}_i || \hat{\bf y}$ and ${\bf k}_s || -\hat{\bf y}$) is not appropriate for the observation of the phonon helicity, as $I_{\rm int}$ vanishes in that case.
We require that the angular momentum of the light be non orthogonal to the phonon angular momentum, like in 
 Fig.~\ref{fig:expsketch}.


Sixth, in  $\alpha-$quartz, the global phase of the Raman tensor for the $E$ phonons is not observable in the Raman intensity: the analogue of $\phi_{f}-\phi_g$ in BaMnSb$_2$ vanishes for $\alpha-$quartz, because the hybridization takes between two phonons of the same irreducible representation ($E$).  In BaMnSb$_2$, hybridization takes place between phonons of different irreducible representations  ($A_1$ and $B_1$). As a result, the complex phases of the corresponding Raman tensors are observable in the Raman intensity (through the term $\phi_f-\phi_g$ in Eq.~(\ref{eq:I_int})), and must be taken into account in the analysis of the phonon helicity.

Seventh, to our knowledge, the phonon hybridization and angular momentum in $\alpha-$quartz are unrelated to the electronic band topology. In BaMnSb$_2$, the phonon angular momentum originates (at least in part) from the valley Chern number of the electrons.


Finally, in Ref.~[\onlinecite{oishi2022selective}], the Raman tensor of the hybridized phonons is inferred from the experimental data.
It is interesting that same result could have been {\em theoretically} anticipated using an appropriate counterpart of Eq.~(\ref{eq:Rpm}) in our paper.
Specifically, the ${\bf q}=0$ Raman tensors for  $E$ phonons  in $\alpha-$quartz are \cite{loudon2001raman}
\begin{equation}
    {\bf R}_{E(x)}=\begin{pmatrix}
        h& 0\\
        0&-h
    \end{pmatrix};
    {\bf R}_{E(y)}=\begin{pmatrix}
        0& -h\\
        -h&0
    \end{pmatrix},
\end{equation}
where $h$ is a complex number. Since these phonons are degenerate at ${\bf q}=0$, we can apply Eq.~(\ref{eq:S_ph}) with $\omega_A=\omega_B$. This sets $\theta_{\bf q} = \pi/2$.
In addition, since the hybridization in $\alpha-$quartz is purely imaginary, we have $\varphi_{\bf q}=\pi/2$.
Consequently, the counterpart of Eq.~(\ref{eq:Rpm}) becomes
\begin{align}
{\bf R}_+ &= \frac{1}{\sqrt{2}}\left({\bf R}_{E(x)}-i {\bf R}_{E(y)}\right) =h\begin{pmatrix}
        1 & i\\
        i & -1
\end{pmatrix} 
\notag\\
{\bf R}_- &= \frac{1}{\sqrt{2}}\left(-{\bf R}_{E(x)}-i {\bf R}_{E(y)}\right) = h\begin{pmatrix}
        -1 & i\\
        i &1
    \end{pmatrix},
    \end{align}
which are identical to the experimentally inferred Raman tensors of Ref.~[\onlinecite{oishi2022selective}].

\section{Summary  and conclusions}
\label{sec:conc}
 Theoretically predicted in Ref.~\cite{hu2021phonon}, the phonon helicity in BaMnSb$_2$ (a nonchiral crystal) awaits experimental validation.
In this work, we have proposed a method to detect the helicity of phonons in BaMnSb$_2$ using Raman spectroscopy.
While Raman-based detection of phonon angular momentum has been recently reported in chiral crystals, our theory could be applied for the detection of phonon angular momentum in a wider class of crystals.

A few general statements can be extracted from our study. First, to detect phonon helicity, the light must have a nonzero angular momentum collinear to the phonon angular momentum.
Second, for a fixed angular momentum of the light, the reversal of the phonon angular momentum results in a change in the Raman intensity.
Likewise, for a fixed angular momentum of the phonon, the reversal of the light's angular momentum leads to a change in the Raman intensity.
With further analysis, the origin of this change can be attributed to the phonon helicity.


More broadly, our results highlight the possibility of detecting signatures of nontrivial electronic band topology by Raman spectroscopy. 
For example, a band inversion in BaMnSb$_2$ might be observable in Raman intensity because of the accompanying reversal in the phonon helicity.
This motivates further theoretical studies to discover new connections between the electronic Berry phase and the Raman spectroscopy. 


\acknowledgements
This work has been financially supported by the Natural Sciences and Engineering Research Council of Canada (Grant No. RGPIN- 2018-05385), and the Fonds de Recherche du Québec Nature et Technologies.
I. G. is grateful to L. Hu, J. Yu and C.-X. Liu for collaboration in Ref.~[\onlinecite{hu2021phonon}], which motivated the present work.
\appendix

\begin{widetext}

\section{Functional integral derivation of the effective phonon action}
\label{sec:app}

The objective of this Appendix is to rederive Eq.~(\ref{eq:S_ph}), which was first obtained in Ref.~[\onlinecite{hu2021phonon}], but using a different approach.
Our calculation will follow the formalism of Ref.~[\onlinecite{rinkel2019influence}], appropriately generalized in order to incorporate phonon hybridization.
Then, we will show that both the real and imaginary parts of the hybridization phonon self-energy $\Sigma_{AB}$ are linear in $q_y$ at long wavelength. While ${\rm Im}\Sigma_{AB}\propto q_y$ was clearly shown in Ref.~[\onlinecite{hu2021phonon}], ${\rm Re}\Sigma_{AB}\propto q_y$ was not clearly stated therein.

\subsection{Preliminaries}
Following Ref.~[\onlinecite{hu2021phonon}], the low-energy free electron Hamiltonian takes form 
\begin{equation}
{\cal H}_{\rm el}^{(0)} = \sum_{\bf k} \Psi^\dagger_{\bf k} h_0({\bf k}) \Psi_{\bf k},
\end{equation}
where $\Psi^\dagger_{\bf k}$ creates a 4-component spinor state at momentum {\bf k}, 
\begin{equation}
\label{eq:h0}
h_0({\bf k}) = s_z\left[v (k_x \tau_x + k_y \tau_z) + m \tau_y\right]
\end{equation}
is an effective Hamiltonian for massive Dirac fermions in two dimensions, 
$m$ is the Dirac mass, $\tau_z$ labels two orbitals and $s_z$ labels two valleys.

The electron-phonon interaction is \cite{hu2021phonon}
\begin{equation}
{\cal H}_{\rm el-ph}(t)=\sum_{{\bf k}, {\bf q}} \Psi^\dagger_{\bf k} \left[ \left(g_0({\bf q}) + g_2({\bf q}) \tau_y s_z + g_3({\bf q}) \tau_z\right) u_A({\bf q},t) + g_1({\bf q} )\tau_x u_B({\bf q},t)\right] \Psi_{{\bf k}+{\bf q}},
\end{equation}
where $t$ is the time and $g_\lambda$ are the matrix elements of the electron-phonon coupling vertex.
The couplings $g_0$, $g_1$ and $g_3$ are pseudo gauge-field like.
The coupling $g_2$ describes phonon-induced dynamical fluctuations in the Dirac mass.
 
\par 
Finally, the electron-electron interaction in the long-wavelength approximation and under the neglect of umklapp processes is 
\begin{equation}
{\cal H}_{\rm el-el} \simeq \frac{1}{2{\cal V}} \sum_{\bf q} U({\bf q}) \rho({\bf q}) \rho(-{\bf q}),
\end{equation}
where ${\cal V}$ is the sample volume, 
\begin{equation}
U({\bf q}) = \frac{e^2}{\epsilon_\infty {\bf q}^2}
\end{equation}
 is the ``bare" Coulomb potential ($\epsilon_\infty$ denotes the screening from high-energy electronic bands that are not included in Eq.~(\ref{eq:h0})) and
\begin{equation}
\rho({\bf q}) = \sum_{\bf k} \Psi^\dagger_{\bf k} \Psi_{\bf k}
\end{equation}
is the electron density operator.

\subsection{Effective action for hybridized phonons}
The imaginary time effective action for phonons, $S_{\rm eff}[u_A,u_B]$, is defined through
\begin{equation}
e^{-S_{\rm eff}[u_A,u_B]}=e^{-S_0[u_A,u_B]} \int D\Psi D\bar{\Psi} \,{\rm exp}\left[-\int d^3 x \bar{\Psi} (\partial_{\bar{t}}-\mu)\Psi - \int d\bar{t} \left({\cal H}_{\rm el}^{(0)}[\Psi,\bar{\Psi}]+{\cal H}_{\rm el-el}[\Psi,\bar{\Psi}]+{\cal H}_{\rm el-ph}[\Psi, \bar{\Psi}, u_A, u_B]\right)\right],
\end{equation}
where $x=({\bf r},\bar{t})$, $\bar{t}$ is the imaginary time, $\mu$ is the chemical potential,
\begin{equation}
S_0[u_A,u_B]\sim\frac{T}{2 {\cal V}} \sum_q \sum_{\lambda=A,B} (q_0^2+\omega_\lambda^2) u_\lambda(q) u_\lambda(-q)
\end{equation}
is the bare phonon action in the absence of Dirac fermions, $T$ is the temperature, $q=({\bf q}, q_0)$, $q_0$ is a Matsubara frequency, $\Psi$ and $\bar{\Psi}$ are Grassmann fields. 

In order to treat the quartic term in fermion fields, we use the Hubbard-Stratonovich transformation,
\begin{equation}
\exp\left[-\frac{T}{2 {\cal V}} \sum_q V({\bf q}) \rho(q) \rho(-q)\right]\propto \int D\varphi \exp\left\{-\frac{T}{\cal V}\sum_q\left[\frac{\epsilon_\infty {\bf q}^2}{2}\varphi(q)\varphi(-q) + i e\varphi(q)\rho(-q)\right]\right\},
\end{equation}
where $\varphi$ is the auxiliary bosonic field associated to the electric potential created by the fluctuations of the electron density.
Accordingly, we have
\begin{equation}
e^{-S_{\rm eff}[u_A, u_B]} \propto e^{-S_0[u_A,u_B]} \int D\varphi D\Psi D\bar{\Psi} \exp\left[-\frac{T}{\cal V} \sum_q \frac{\epsilon_\infty {\bf q}^2}{2}\varphi(q)\varphi(-q) + \int d^3x \bar{\Psi}(x) G^{-1}(x,x)\Psi(x)\right],
\end{equation}
where
\begin{equation}
G^{-1}(x,x)=G_0^{-1}(x,x)-V(x)
\end{equation}
is the inverse of the full fermionic Green's function, and the Green's function for free Dirac fermions  is defined as 
$\hbar [-\partial_{\bar t} + \mu - v(-i\tau_x \partial_x - i \tau_z\partial_y) s_z-m\tau_y s_z]G_0(x,x')=\delta(x-x')$.
Additionally, 
\begin{equation}
V(x)= i e\varphi(x) + \sum_{\bf q} e^{i {\bf q}\cdot{\bf r}} \left[ \left(g_0({\bf q}) + g_2({\bf q}) \tau_y s_z + g_3({\bf q}) \tau_z\right) u_A(q) + g_1({\bf q} )\tau_x u_B(q)\right]
\end{equation}
is the correction due to electron-phonon and electron-electron interactions.

The integral over the fermionic fields is now gaussian and can be done exactly. 
Assuming the absence of lattice instabilities and electronic ordered states, we can expand the resulting expression to lowest order in powers of $u_\lambda$ and $\varphi$. 
The outcome reads
\begin{align}
e^{-S_{\rm eff}[u_A,u_B]} \propto e^{-S_0[u_A,u_B]} \int D\varphi \exp&\left\{-\frac{T}{\cal V}\sum_q \frac{\epsilon_\infty {\bf q}^2}{2}\varphi(q)\varphi(-q) 
- \frac{\hbar}{2} \int d^3 x d^3 x' {\rm tr}\left[G_0(x,x') V(x') G_0(x',x) V(x)\right]\right\},
\end{align}
where tr stands for trace in the $(\tau, s)$ space.
Because of translational symmetry, 
\begin{equation}
\int d^3 x d^3 x' {\rm tr}\left[G_0(x,x') V(x') G_0(x',x) V(x)\right] = \frac{T^2}{{\cal V}^2} \sum_q \sum_k {\rm tr}\left[G_0(k) V(q) G_0(k+q) V(-q)\right],
\end{equation}
where 
\begin{equation}
V(q) =  i e\varphi(q) +   \left(g_0({\bf q}) + g_2({\bf q}) \tau_y s_z + g_3({\bf q}) \tau_z\right) u_A(q) + g_1({\bf q} )\tau_x u_B(q).
\end{equation}
Then,
\begin{align}
\label{eq:seff1}
e^{-S_{\rm eff}[u_A,u_B]} \propto e^{-S_0[u_A,u_B]} \int D\varphi \exp &\left\{-\frac{T}{2\cal V}\sum_q\epsilon_\infty {\bf q}^2\varphi(q)\varphi(-q) 
- \frac{T}{2 \cal V} \sum_q \hbar\frac{T}{\cal V}\sum_k {\rm tr}\left[G_0(k) V(q) G_0(k+q) V(-q)\right]\right\}.
\end{align}
Let us define
\begin{align}
\Pi_{i j}^{(0)}(q) &= \hbar\frac{T}{\cal V} \sum_k {\rm tr}\left[G_0(k) \tau_i G_0(k+q) \tau_j\right]\nonumber\\
\Pi_{i j}^{(z)}(q) &= \hbar\frac{T}{\cal V} \sum_k {\rm tr}\left[G_0(k) \tau_i G_0(k+q) \tau_j s_z\right].
\label{eq:apdfpt}
\end{align}
Note that, in the expression for $\Pi_{i j}^{(z)}$, it does not matter where we put the $s_z$ matrix, because $G_0$ commutes with $s_z$.
With these definitions, we can write
\begin{align}
\epsilon_\infty {\bf q}^2\varphi(q)\varphi(-q)+\hbar\frac{T}{\cal V}\sum_k {\rm tr}\left[G_0(k) V(q) G_0(k+q) V(-q)\right]= i e \left[\alpha(-q) \varphi(q)+\alpha(q) \varphi(-q)\right]+e^2 \beta(q) \varphi(q)\varphi(-q) + \delta(q) 
\end{align}
with
\begin{align}
\alpha(-q) & = g_1(-{\bf q}) u_B(-q) \Pi_{01}^{(0)}(q) +g_0(-{\bf q}) u_A(-q) \Pi_{00}^{(0)}(q)\nonumber\\
&+g_2(-{\bf q}) u_A(-q) \Pi_{02}^{(z)}(q) +g_3(-{\bf q}) u_A(-q)  \Pi_{03}^{(0)}(q) \nonumber\\
\beta(q)&=\epsilon_\infty {\bf q}^2/e^2-\Pi_{00}^{(0)}(q)=\beta(-q)\nonumber\\
\delta(q) &= u_A(q) u_A(-q) \left\{g_0(-{\bf q}) g_0({\bf q})  \Pi_{00}^{(0)}(q) +g_2(-{\bf q}) g_2({\bf q})  \Pi_{22}^{(0)}(q) +g_3(-{\bf q}) g_3({\bf q})  \Pi_{33}^{(0)}(q)\right.\nonumber\\
 & \left. +\left[g_2(-{\bf q}) g_0({\bf q})  \Pi_{02}^{(z)}(q) +g_3(-{\bf q}) g_0({\bf q})  \Pi_{03}^{(0)}(q)+g_2(-{\bf q}) g_3({\bf q}) \Pi_{32}^{(z)}(q) + (q\leftrightarrow -q) \right]\right\}\nonumber\\
 &+u_B(-q) u_B(q) g_1(-{\bf q}) g_1({\bf q})  \Pi_{11}^{(0)}(q)\nonumber\\
&+u_A(q) u_B(-q) \left\{ g_1(-{\bf q}) g_0({\bf q})  \Pi_{01}^{(0)}(q)+g_1(-{\bf q}) g_2({\bf q}) \Pi_{21}^{(z)}(q)+ g_1(-{\bf q}) g_3({\bf q})  \Pi_{31}^{(0)}(q)\right\}\nonumber\\
&+u_A(-q) u_B(q) \left\{ g_1({\bf q}) g_0({-\bf q})  \Pi_{01}^{(0)}(-q) +g_1({\bf q}) g_2(-{\bf q}) \Pi_{21}^{(z)}(-q)+ g_1({\bf q}) g_3(-{\bf q}) \Pi_{31}^{(0)}(-q)\right\}\nonumber\\
&=\delta(-q)
 \end{align}
Next, we complete a square in order to decouple the $\varphi$ field from the phonon fields $u_\lambda$:
\begin{equation}
\epsilon_\infty {\bf q}^2\varphi(q)\varphi(-q)+\hbar\frac{T}{\cal V}\sum_k {\rm tr}\left[G_0(k) V(q) G_0(k+q) V(-q)\right]=e^2 \beta(q) \left[\varphi(q)+\frac{i}{e}\frac{\alpha(q)}{\beta(q)}\right]\left[\varphi(-q)+\frac{i}{e}\frac{\alpha(-q)}{\beta(-q)}\right]+\frac{\alpha(q) \alpha(-q)}{\beta(q)}+\delta(q),
\end{equation}
where we have exploited the fact that $\beta(q)=\beta(-q)$.
Now, because $\varphi$ is being integrated in Eq.~(\ref{eq:seff1}), we can make a shift
\begin{equation}
\varphi(q) \to \varphi(q) -\frac{i}{e}\frac{\alpha(q)}{\beta(q)}
\end{equation}
without changing the value of the integral.
This results in
\begin{equation}
\label{eq:seff2}
e^{-S_{\rm eff}[u_A,u_B]} \propto e^{-S_0[u_A,u_B]-\delta S[u_A,u_B]} \int D\varphi \exp\left[-\frac{T}{2\cal V}\sum_q e^2 \beta(q)\varphi(q)\varphi(-q) \right],
\end{equation}
where
\begin{equation}
\label{eq:dS}
\delta S[u_A,u_B]=\frac{T}{2\cal V} \sum_q \left[\delta(q)+\frac{\alpha(q) \alpha(-q)}{\beta(q)}\right]
\end{equation}
depends only on the phonon fields. 
In Eq.~(\ref{eq:seff2}), we notice that the coefficient of $\varphi(q)\varphi(-q)$ has become 
\begin{equation}
e^2 \beta(q) = \epsilon_\infty {\bf q}^2 \left[1-U({\bf q}) \Pi_{00}^{(0)}(q)\right] \equiv \epsilon_\infty \epsilon(q) {\bf q}^2. 
\end{equation}
This result is standard: the action for the electromagnetic field is renormalized by screening from conduction electrons, which amounts to replacing $\epsilon_\infty$ by $\epsilon_\infty \epsilon(q)$ in the bare electromagnetic action.
Here, $\epsilon(q)$ is the RPA contribution from low-energy Dirac fermions to the dielectric function.
In Eq.~(\ref{eq:dS}), the first term is the contribution from electron-phonon interactions to the phonon dynamics, in the absence of electron-electron interactions. 
The influence of electron-electron interactions in phonon dynamics is contained in the last term of Eq.~(\ref{eq:dS}); note that $\alpha(q) \alpha(-q)/\beta(q)$ vanishes when $e\to 0$. 

Now that we have decoupled the phonon fields from the electromagnetic fields, the effective action for phonons reads (modulo terms that do not depend on phonon fields and thus do not contribute to phonon dynamics)
\begin{equation}
S_{\rm eff}[u_A,u_B] = S_0[u_A,u_B]+\delta S[u_A,u_B].
\end{equation}
Let us consider for a moment the simple case (treated e.g. in Ref.~[\onlinecite{mahan2013many}]) in which we consider a single phonon mode (say, the A mode), which couples to electrons only through $g_0$ (scalar potential). 
Then, we obtain
\begin{equation}
\delta(q)+\frac{\alpha(q) \alpha(-q)}{\beta(q)} =\frac{|g_0({\bf q})|^2}{\epsilon(q)}  \Pi_{00}^{(0)}(q) |u_A(q)|^2,
\end{equation}
where we have used $g_i(-{\bf q}) = g_i({\bf q})^*$ for $i=0$.
 This is a familiar result: the effect of electron-electron interactions is to screen the electron-phonon vertex through
 \begin{equation}
 |g_0({\bf q})|^2 \to \frac{|g_0({\bf q})|^2}{\epsilon(q)}.
 \end{equation}
For an electronic insulator, $\epsilon(q)$ is roughly a constant in the long wavelength limit (which comes from the interband contributions to $\Pi_{00}^{(0)}$)
 
Now that we have verified that the structure of Eq.~(\ref{eq:dS}) makes sense in simple situations, let us return to our problem of interest.
The full effective action for the phonons can be written as
\begin{equation}
\label{eq:seffAB}
S_{\rm eff}[u_A,u_B] = \frac{T}{2\cal V} \sum_q \left(u_A(-q), u_B(-q)\right) \left(\begin{array}{cc} q_0^2 + \omega_A({\bf q})^2 + \Sigma_{AA}(q) & \Sigma_{AB}(q) \\ \Sigma_{BA}(q) &q_0^2 + \omega_B({\bf q})^2+\Sigma_{BB}(q)\end{array}\right) \left(\begin{array}{c}u_A(q) \\ u_B(q) \end{array}\right),
\end{equation}
where 
\begin{align}
\Sigma_{AA}(q)&=|g_0({\bf q})|^2  \Pi_{00}^{(0)}(q) +|g_2({\bf q})|^2  \Pi_{22}^{(0)}(q) +|g_3({\bf q})|^2  \Pi_{33}^{(0)}(q)\nonumber\\
 & +\left[g_2(-{\bf q}) g_0({\bf q})  \Pi_{02}^{(z)}(q) +g_3(-{\bf q}) g_0({\bf q})  \Pi_{03}^{(0)}(q)+g_2(-{\bf q}) g_3({\bf q}) \Pi_{32}^{(z)}(q) + (q\leftrightarrow -q) \right]\nonumber\\
 &+\frac{U({\bf q})}{\epsilon(q)}\left[ |g_0({\bf q})|^2 \left(\Pi_{00}^{(0)}(q)\right)^2+ |g_2({\bf q})|^2 \Pi^{(z)}_{02}(q) \Pi^{(z)}_{02}(-q)+|g_3({\bf q})|^2 \Pi^{(0)}_{03}(q) \Pi^{(z)}_{03}(-q) \right]\nonumber\\
  &+\frac{U({\bf q})}{\epsilon(q)}\left[ g_0(-{\bf q}) g_2({\bf q}) \Pi_{00}^{(0)}(q) \Pi_{02}^{(z)}(-q)+ g_0(-{\bf q}) g_3({\bf q}) \Pi_{00}^{(0)}(q) \Pi_{03}^{(0)}(-q)+g_2(-{\bf q}) g_3({\bf q}) \Pi_{02}^{(z)}(q) \Pi_{03}^{(0)}(-q) + (q\leftrightarrow -q)\right]\nonumber\\
  &=\Sigma_{AA}(-q),\nonumber\\
  \Sigma_{BB}(q)&=  |g_1({\bf q})|^2  \Pi_{11}^{(0)}(q) + \frac{U({\bf q})}{\epsilon(q)} |g_1({\bf q})|^2  \left(\Pi_{11}^{(0)}(q) \right)^2=\Sigma_{BB}(-q),\nonumber\\
  \Sigma_{AB}(q)&= g_1({\bf q}) g_0({-\bf q})  \Pi_{01}^{(0)}(-q) +g_1({\bf q}) g_2(-{\bf q}) \Pi_{21}^{(z)}(-q)+ g_1({\bf q}) g_3(-{\bf q}) \Pi_{31}^{(0)}(-q)\nonumber\\
  &+\frac{U({\bf q})}{\epsilon(q)}\left[ g_0(-{\bf q}) g_1({\bf q}) \Pi_{00}^{(0)} (q) \Pi_{01}^{(0)}(-q)+g_2(-{\bf q}) g_1({\bf q}) \Pi_{02}^{(z)} (q) \Pi_{01}^{(0)}(-q)+g_3(-{\bf q}) g_1({\bf q}) \Pi_{03}^{(0)} (q) \Pi_{01}^{(0)}(-q)\right]\nonumber\\
  &=\Sigma_{BA}(-q).
 \end{align}
 In Eq.~(\ref{eq:S_ph}), we have absorbed the diagonal self-energies $\Sigma_{AA}$ and $\Sigma_{BB}$ into  $\omega_A^2$ and $\omega_B^2$.

\subsection{Calculation of $\Sigma_{AB}$ at long wavelengths}
In this section, we will study the $q_y$ dependence of the both real and imaginary part of $\Sigma_{AB}$ at long wavelengths.
To that end, we need to calculate $\Pi^{(0)}_{01}$,$\Pi^{(0)}_{31}$ and $\Pi^{(z)}_{21}$ (for simplicity, we omit the effect of $U({\bf q})$ in this discussion).
Assuming $\mu=0$ (Fermi energy inside the gap of the insulator), we have 
\begin{equation}
G_0^{-1}(k)=i \omega_n - h_0({\bf k}).
\label{eq:apgrnfnc}
\end{equation}
From Eq.~(\ref{eq:apdfpt}), we get (in the zero-temperature limit)
\begin{align}
\Pi^{(0)}_{01}(q) &=\frac{1}{(2\pi)^3}\sum_{s=\pm 1} \int d^2 k \frac{-i q_0 ( E_{\bf k} (s k_x + s q_x)-E_{{\bf k}+{\bf q}} s k_x) + 
 i  m  q_y(E_{\bf k} + E_{{\bf k}+{\bf q}}) }{E_{\bf k} E_{{\bf k}+{\bf q}} ((E_{\bf k} + E_{{\bf k}+{\bf q}})^2 + q_0^2)} \notag\\
 &= 
 \frac{2}{(2\pi)^3}\int d^2 k \frac{ 
 i  m  q_y(E_{\bf k} + E_{{\bf k}+{\bf q}}) }{E_{\bf k} E_{{\bf k}+{\bf q}} ((E_{\bf k} + E_{{\bf k}+{\bf q}})^2 + q_0^2)}
 ,
\end{align}
where $E_{\bf k}=\sqrt{v^2 (k_x^2+k_y^2)+m^2}$ (we take $v\equiv 1$ hereafter).
Thus, to leading order in the long-wavelength approximation, 
\begin{equation}
\Pi^{(0)}_{01}(q) \simeq \frac{4}{(2\pi)^3} \int d^2 k \frac{ 
 i  m  q_y  }{E_{\bf k}  (4 E_{\bf k}^2 + q_0^2)}.
 \end{equation}
Applying the analytical continuation $q_0 \rightarrow -i\omega+\eta$ and utilizing the Plemelj-Shokhotski formula
\begin{equation}
\frac{1}{x\pm i \eta}=\mathcal P \frac{1}{x} \mp i\pi \delta(x),
\label{eq:applemelj}
\end{equation} 
we get
\begin{align}
\label{eq:Pi01}
&{\rm Im}\, \Pi^{(0)}_{01}({\bf q},\omega)\simeq \frac{4}{(2\pi)^3} \int d^2 k \frac{ 
 i  m  q_y  }{E_{\bf k}  (4 E_{\bf k}^2 - \omega^2)}\notag\\
&{\rm Re}\, \Pi^{(0)}_{01}({\bf q},\omega) \simeq -\frac{4}{(2\pi)^3} \int d^2 k \frac{ 
 m  q_y {\rm sgn}(\omega)  }{E_{\bf k} }\delta(4 E_{\bf k}^2 - \omega^2).
\end{align}
Thus, both ${\rm Im}\, \Pi^{(0)}_{01}({\bf q},\omega)$ and ${\rm Re}\, \Pi^{(0)}_{01}({\bf q},\omega)$ are linear in $q_y$ at long wavelengths. 
In addition, ${\rm Im}\, \Pi^{(0)}_{01}({\bf q},\omega)$ is even in $\omega$, while ${\rm Re}\, \Pi^{(0)}_{01}({\bf q},\omega)$ is odd in $\omega$.

At zero temperature and in the absence of disorder, the real part of the hybridization self-energy is nonzero if the phonon frequency $\omega$ exceeds the gap of the insulator ($ 2|m|$).
In the presence of disorder, the Dirac delta function in Eq.~(\ref{eq:Pi01}) is broadened into a Lorentzian, and therefore the real part of $\Sigma_{AB}$ is nonzero even when the phonon frequency $\omega$ is lower than the gap of the insulator.

Expressions for $\Pi^{(0)}_{31}$ and $\Pi^{(z)}_{21}$ can be similarly obtained. 
It follows that both are odd functions of $q_y$, and in both cases the real and imaginary parts have opposite parity under $\omega\to-\omega$.
It is worth to mention that the contribution of $\Pi^{(z)}_{21}$, associated to the coupling $g_2$,  makes the $2\times 2$ matrix in Eq.~(\ref{eq:S_ph}) non hermitian for real frequencies.
In the main text, we have disregarded this term for simplicity and have treated the contributions from $\Pi^{(0)}_{01}$ and $\Pi^{(0)}_{31}$, which are associated to the pseudo gauge-field like couplings $g_0$, $g_1$ and $g_3$.

 \section{Some relations for the Raman tensor elements}
\label{sec:appB}

In Eq.~(\ref{eq:ramantensor}), as well as in preceding works \cite{ribeiro2015unusual, resende2020origin, han2022complex}, Raman tensors at zero wave vector are considered to be complex symmetric matrices.
The purpose of this appendix is to assess the conditions under which Raman tensors in Eq. (\ref{eq:ramantensor}) are symmetric even when the coefficients are complex. To that end, we follow the formalism of Ref. \cite{loudon1963theory}. 

There are two assumptions in Ref.~\cite{loudon1963theory}, which consist of neglecting the finite lifetime of the electrons and restricting to real electronic wave functions. Yet, the finite lifetime of electrons can lead to complex phases of Raman tensor elements, and in the presence of spin-orbit interactions the electronic wave functions need not be real in spite of time-reversal symmetry.
If we relax those two assumptions,  the matrix elements of the zero-temperature Raman tensor from Ref.~\cite{loudon1963theory} take the form 
\begin{equation}
\begin{split}
    R^i_{12}(-\omega_1,\omega_2,\omega_0;\Gamma)=\frac{1}{V} \sum_{\alpha \beta}& \Bigg \{ \frac{p^2_{0\beta}p^1_{\beta \alpha} \Xi^i_{\alpha 0}}{(\omega_{\beta}+\omega_0-\omega_1+i\Gamma_{\beta})(\omega_{\alpha}+\omega_0+i\Gamma_{\alpha})}+    
   \frac{p^1_{0\beta}p^2_{\beta \alpha} \Xi^i_{\alpha 0}}{(\omega_{\beta}+\omega_0+\omega_2+i\Gamma_{\beta})(\omega_{\alpha}+\omega_0+i\Gamma_{\alpha})} + \\
   &\frac{p^2_{0\beta} \Xi^i_{\beta \alpha} p^1_{ \alpha 0}}{(\omega_{\beta}+\omega_0-\omega_1+i\Gamma_{\beta})(\omega_{\alpha}-\omega_1+i\Gamma_{\alpha})}+    
   \frac{p^1_{0\beta} \Xi^i_{\beta \alpha} p^2_{\alpha 0}}{(\omega_{\beta}+\omega_0+\omega_2+i\Gamma_{\beta})(\omega_{\alpha}+\omega_2+i\Gamma_{\alpha})} + \\
   &\frac{ \Xi^i_{0 \beta} p^2_{\beta \alpha}p^1_{\alpha 0}}{(\omega_{\beta}+\omega_2-\omega_1+i\Gamma_{\beta})(\omega_{\alpha}-\omega_1+i\Gamma_{\alpha})}+    
   \frac{ \Xi^i_{0 \beta} p^1_{\beta \alpha}p^2_{\alpha 0}}{(\omega_{\beta}+\omega_2-\omega_1+i\Gamma_{\beta})(\omega_{\alpha}+\omega_2+i\Gamma_{\alpha})}  \Bigg \} 
   \label{eq:thirdorderperturb},
\end{split}
\end{equation}
where we preserved the notation of Ref. \cite{loudon1963theory} except for the addition of an inverse lifetime factor $\Gamma$ for the intermediate electronic many-body excited states $\alpha$ and $\beta$.  Here,  $1$ and $2$ ($\omega_1$ and $\omega_2$)  denote the polarizations (frequencies) of the incoming and scattered light, respectively. Likewise, $i\in\{x,y,z\}$ ($\omega_0$) denotes the polarization direction (the frequency) of the phonon. 
Note that, in the main text, the phonon frequency is instead denoted as $\omega_{{\bf q},\lambda}$.
For a Stokes process, we can effectively take $\omega_1=\omega_2+\omega_0$.
Also, $\Xi_{\alpha\beta}^i$ and $p_{\alpha\beta}^i$ are the matrix elements of the electron-phonon and electron-photon interactions. 
For simplicity of notation, we take $\Gamma_\alpha=\Gamma$ for all $\alpha$.

By inspection of Eq. (\ref{eq:thirdorderperturb}), we can obtain
\begin{align}
     &R^i_{12}(-\omega_1,\omega_2,\omega_0;\Gamma)= R^{i*}_{12}(\omega_1,-\omega_2,-\omega_0,-\Gamma), 
     \label{eq:ramanrelations1}
     \\
      &R^i_{12}(-\omega_1,\omega_2,\omega_0;\Gamma)= R^i_{21}(\omega_2,-\omega_1,\omega_0;\Gamma).
      \label{eq:ramanrelations2}
\end{align}
Combining Eq. (\ref{eq:ramanrelations1}) and Eq. (\ref{eq:ramanrelations2}), we have 
\begin{equation}
     R^i_{12}(-\omega_1,\omega_2,\omega_0;\Gamma)= R^{i*}_{21}(-\omega_2,\omega_1,-\omega_0;-\Gamma).
\end{equation}
If we are far from the resonance (in practice, if we can neglect $\omega_0$ in Eq.~(\ref{eq:thirdorderperturb})), we obtain
\begin{equation}
     R^i_{12}(-\omega_1,\omega_1;\Gamma)= R^{i*}_{21}(-\omega_1,\omega_1,;-\Gamma).
     \label{eq:symmetricrel}
\end{equation}
This equation indicates that the Raman tensor is a complex symmetric provided that (i) we are far from resonance, and (ii) the modulus (phase) of the matrix elements are even (odd) functions of $\Gamma$.
In order to prove the second statement, 
we apply Eq. (\ref{eq:symmetricrel}) to the diagonal matrix elements of the Raman tensor ${\bf R}_A$  in Eq. (\ref{eq:ramantensor}):
\begin{align}
\label{eq:Rxx}
R^x_{xx}(-\omega_1,\omega_1;\Gamma)= R^{x*}_{xx}(-\omega_1,\omega_1;-\Gamma) \rightarrow|a(\Gamma)|e^{i\phi(\Gamma)}=|a(-\Gamma)|e^{-i\phi(-\Gamma)},
\end{align}
where $R^{x}$ refers to $R_{A}$ in the main text (the $A$ phonon therein is polarized along $x$). It follows from Eq.~(\ref{eq:Rxx}) that 
the modulus of a Raman tensor element must be an even function of $\Gamma$, while the phase must be an odd function of $\Gamma$.
Assuming reasonably that the same property holds for the matrix elements of ${\bf R}_B$, we conclude that ${\bf R}_B$ will be a complex symmetric matrix.


\section{Raman intensity in BaMnSb$_2$ under linearly polarized light}
\label{sec:appC}

In the main text, we have considered circularly polarized incident and scattered lights for the detection of the phonon helicity.
In this Appendix we justify why we cannot perform this detection using a linearly polarized light.

For general linear polarizations of the incident and scattered lights, we have
\begin{equation}
I \propto \left | \hat{\bf e}_i^\dagger\cdot {\bf R}_\lambda \cdot \hat{\bf e}_s\right |^2 =\left |(ax x^\prime+ b y y^\prime+ c z z^\prime) \cos \left( \frac{\theta}{2} \right)+d e^{-i \varphi_{\bf q}}(x y^\prime+ y x^\prime) \sin \left( \frac{\theta}{2} \right) \right|^2,
\label{eq:ap:polarization}
\end{equation}
where $(x, y, z)$, and $(x^\prime, y^\prime, z^\prime)$ are the polarizations of incoming light and scattered light respectively, all components being real.  The term of interest (containing $\varphi_{\bf q}$) is the crossed term in Eq.~(\ref{eq:ap:polarization}), which we have called ``interference term" in the main text.
Since this term does not require $z$ and $z'$, let us herein take $z=z'=0$ for simplicity.
Then, the interference term takes the form
\begin{equation}
I_{\rm int} \propto {\rm Re} \left( e^{-i \varphi_{\bf q}}(ax x^\prime+ b y y^\prime)(dx y^\prime+d y x^\prime)  \right).
\label{eq:ap:interference}
\end{equation} 
At first glance, it appears that $I_{\rm int}$ is odd under $q_y\to -q_y$, since $\varphi_{\bf q} \to \varphi_{\bf q} + \pi$ under the transformation. 
Then, one can once again extract $I_{\rm int}$ experimentally by comparing the Raman intensities at $\pm q_y$. 
Yet, as we explain next, it turns out that $I_{\rm int}$ is even under $q_y\to -q_y$.
The reason is that the polarization vector of the light must also be changed when $q_y\to -q_y$.
To see this, we recall the condition from Maxwell's equations:
\begin{align}
\label{eq:Max}
&{\bf k}_i. \hat{\bf e}_i=k^i_{x}x+k^i_{y}y=0, \notag\\
&{\bf k}_s. \hat{\bf e}_s=k^s_{x}x^\prime+k^s_{y}y^\prime=0.
\end{align}
If $k^i_{y}$ and $k^s_{y}$ both change sign, we have $q_y \rightarrow -q_y$. 
But, in order to still satisfy Eq.~(\ref{eq:Max}), 
either $x$ or $y$ must change sign, and similarly either $x^\prime$ or $y^\prime$ must change sign. 
This gives four possibilities. In all four of them, quick inspection of Eq.~(\ref{eq:ap:interference}) shows that $I_{\rm int}$ remains unchanged under $q_y\to -q_y$. 
Accordingly, linearly polarized light is not convenient to separate out the interference term containing the information about the phonon helicity. The circular polarization configurations discussed in the main text do not have this problem.

\end{widetext}

\bibliography{whole}{}
\bibliographystyle{apsrev4-1}

\end{document}